\documentclass[twocolumn,showpacs,amsmath,amssymb,superscriptaddress]{revtex4-1}
\usepackage{graphicx}% Include figure files
\usepackage{dcolumn}% Align table columns on decimal point
\usepackage{bm}% bold math
\usepackage[abs]{overpic}
\usepackage{here}
\usepackage{color}
\usepackage{url}
\def\Vec#1{\bm{#1}}

\begin{document}

%\preprint{}

\title{
Reduced-Shifted Conjugate-Gradient Method for a Green's Function: \\
Efficient Numerical Approach in a Nano-structured Superconductor
%Efficient numerical solver for a nano-structured superconductor:
%\\
%reduced-shifted Krylov Bogoliubov-de Gennes solver
}
\author{Yuki Nagai}
\affiliation{CCSE, Japan  Atomic Energy Agency, 178-4-4, Wakashiba, Kashiwa, Chiba, 277-0871, Japan}
\author{Yasushi Shinohara}
\affiliation{Photon Science Center, School of Engineering, The University of Tokyo, 7-3-1 Hongo, Bunkyo-ku, Tokyo, 113-8656, Japan}
%\affiliation{Photon Science Center, School of Engineering, the University of Tokyo}
\author{Yasunori Futamura}
\affiliation{Department of Computer Science, University of Tsukuba, Tsukuba, Ibaraki 305-8573, Japan
}
\author{Tetsuya Sakurai}
\affiliation{Department of Computer Science, University of Tsukuba, Tsukuba, Ibaraki 305-8573, Japan
}

\date{\today}% It is always \today, today,
             
\begin{abstract}
We propose the reduced-shifted Conjugate-Gradient (RSCG) method, which is numerically efficient to calculate a matrix element of a Green's function defined as a resolvent of a Hamiltonian operator, by solving linear equations with a desired accuracy.  
This method does not calculate solution vectors of linear equations but does directly calculate a matrix element of the resolvent. 
The matrix elements with different frequencies are simultaneously obtained. 
Thus, it is easy to calculate the exception value expressed as a Matsubara summation of these elements.
To illustrate a power of our method, we choose a nano-structured superconducting system with a mean-field Bogoliubov-de Gennes (BdG) approach. 
This method allows us to treat with the system with the fabrication potential, where one can not effectively use the kernel-polynomial-based method.
We consider the $d$-wave nano-island superconductor by simultaneously solving the linear equations with a large number ($\sim 50000$) of Matsubara frequencies. 
\end{abstract}

\pacs{
%74.81.-g, %	Inhomogeneous superconductors and superconducting systems, including electronic inhomogeneities
%74.78.Na %Mesoscopic and nanoscale systems
%74.20.Rp, %Pairing symmetries (other than s-wave)
%74.25.Op, %Mixed states, critical fields, and surface sheaths
%74.81.-g	%Inhomogeneous superconductors and superconducting systems, including electronic inhomogeneities
%74.25.Bt  %Thermodynamic properties
}
\maketitle

\section{Introduction}

A Green's function in a quantum field theory plays a key role to describe elemental excitations in both high-energy and condensed matter physics. 
Calculating a Green's function without knowing all eigenfunctions of a many-body Hamiltonian is one of strong theoretical tools to investigate physical systems.
A exception value is described by a summation of Matsubara Green's functions on an imaginary frequency axis. 
A finite order parameter, which is one of exception values, causes a phase transition with a spontaneous symmetry breaking.  
In a topological phase without a spontaneous symmetry breaking\cite{Thouless, Hasan}, the corresponding bound states at a boundary is calculated by a Green's function. 
The kernel-polynomial method (KPM) is one of efficient approaches to calculate a Green's function\cite{Weisse}. 
We note, however, that this method can calculate a Green's function only near the real-frequency axis. 

One of the possible approaches to calculate a Green's function is to solve linear equations, since 
a Green's function is defined as a resolvent of a Hamiltonian operator $G(z) \equiv \left[z I - H \right]^{-1}$ with a complex frequency $z$. 
Each matrix element of the resolvent is connected to a corresponding excitation. 
It seems unreasonable that a massively amount of linear equations at each frequency to calculate the matrix elements by a brute-force approach.

We propose that, with the use of our novel Conjugate-Gradient (CG)-based method, solving linear equations becomes effective in terms of calculations of exception values. 
Our method, which we call the reduced shifted CG (RSCG) method, does not calculate solution vectors of linear equations but does directly calculate a matrix element of the resolvent.
On the basis of the shifted CG-type method\cite{Takayama,Ohno}, we develop the algorithm which updates {\it reduced} quantities realized by matrix operations to the solution vectors  in the CG iteration loops, as shown in Table~\ref{table:RSCG}.

To illustrate power of our method, we choose a nano-structured superconducting system with a mean-field Bogoliubov-de Gennes (BdG) approach. 
There are several reasons as follows. 
Firstly, the mean-field description of inhomogeneous superconductivity through the BdG equations has been highly successful to find unconventional physical phenomena and the BdG equations in real space have been used in a large number of situations where translational symmetry is broken. 
Secondly, the physically important eigenvalues of the BdG Hamiltonian are located around a center of the eigenvalue distribution and the self-consistent calculation needs  the half of eigenvalues and eigenvectors. 
This means that the numerical methods which can calculate limited number of small eigenvalues, such as the Lanczos method, can not be adopted. 
Thirdly, the KPM, which is another Green's function-based method, has been used in many papers\cite{Covaci,NagaiJPSJ,Januszewski,Yoshizawa}. 
We show that the RSCG method allows us to treat the system  where one can not use the KPM effectively 
due to a fabrication-potential-induced broad distribution of eigenvalues.
Finally, an accurate self-consistent calculation has a great demand, since superconductivity has multi length scales (i.e. an inverse of a Fermi-wave length ($\sim$ \AA), a superconducting coherence length ($\sim$ 10-100 nm), and a magnetic field penetration length ($\sim \mu$m)). 
This method calculates a mean field with a desired accuracy.

This paper is organized as follows. 
In Sec.~\ref{sec:general}, we propose the general formulation of the RSCG method. 
We firstly introduce the CG and shifted CG methods to describe the RSCG method. 
In Sec.~\ref{sec:bdg}, we focus on the formulation for the BdG theory of superconductivity. 
We show the self-consistent calculation scheme with the use of the RSCG method. 
In Sec.~\ref{sec:demo}, we illustrate a power of the RSCG method in the $d$-wave nano island and $s$-wave vortex lattice. 
We simultaneously solve the linear equations with a large number ($\sim 50000$) of Matsubara frequencies. 
In Sec.~\ref{sec:tech}, we refer to technical remarks to describe the advantages of the RSCG method. 
The conclusion is given in Sec.~\ref{sec:con}.

\section{Formulation of the reduced-shifted conjugate-gradient method} \label{sec:general}
%\subsection{Conjugate-gradient-based method for Green's functions}
%\subsection{Linear equation Conjugate-gradient-based method for Green's functions}
An important quantity in the quantum field theory is a matrix element of the resolvent of the Hamiltonian operator $H$ called Green function: 
\begin{align}
G_{\alpha \beta}(z) &\equiv \langle \alpha | \left[z I - H \right]^{-1} | \beta \rangle,
\end{align}
where a vector $|\alpha \rangle$ is characterized by an index $\alpha$ in the Hilbert space. 
For example, a zero-temperature general dynamical correlation function with given two operators $A$ and $B$ can be defined through 
\begin{align}
\langle A;B \rangle_{\omega} \equiv \lim_{\epsilon \rightarrow 0} \langle 0| A [(\omega + i \epsilon)I - H]^{-1} B |0 \rangle,
\end{align}
where $|0 \rangle$ is a many-body ground state\cite{Weisse}. 
By solving the linear equation defined as 
\begin{align}
 \left[z I - H \right] |x_{\beta}(z) \rangle &= |\beta \rangle,
\end{align}
the matrix element is expressed as 
\begin{align}
G_{\alpha \beta}(z) &= \langle \alpha | x_{\beta}(z) \rangle, \label{eq:gnm}
\end{align}
without explicit diagonalization of the Hamiltonian.
Representation of the problem is chosen as finite dimension for numerical calculation eventually.
For large and sparse problems, iterative linear solvers are favorable rather than direct solver such as LU-decomposition due to avoid cubic scale increase of calculation cost.

Throughout this section, we usually use linear algebraic expression to avoid misinterpretation.
In order to focus on reason why RSCG works well for our purpose, we go through only essence CG and shifted CG solvers.
Rigorous proof can be found in \cite{SaadCG} and \cite{FrommerSBiCG}.

%It is well known that this linear equation can be solved by the CG-based method \cite{Sup}. 
%We briefly describe the Conjugate Gradient (CG) method and shifted CG method as follows. 

%In this method, the solution vector $|x_{m}(z) \rangle$ is constructed within the Krylov subspace (KS) 
%$K_{n}(z I - H,|m \rangle) \equiv {\rm span} \: \{
%|m \rangle, [z I - H] |m \rangle,\cdots, [z I - H]^{n-1} |m \rangle
% \}$ at the $n$-th iteration. 
%There are various kinds of Krylov-subspace-based methods, such as the conjugate orthogonal conjugate gradient (COCG) method for a complex symmetric matrix, and 
%the Biconjugate gradient (BiCG) method for a non-hermitian matrix. 
%It seems, however, that solving linear equations to obtain $G_{nm}(z)$ has a critical problem, since 
%one has to solve the different linear equations when changing $z$. 
%To overcome this problem, the shifted COCG method has been proposed\cite{Takayama}.
%We describe the Conjugate Gradient (CG) method and shifted CG method as follows. 

\subsection{Conjugate Gradient (CG) method and Krylov subspace}
%Here we introduce the formulation of the conjugate gradient (CG) method.
We consider a linear equation expressed as 
\begin{align}
A x = b
\end{align}
with $x,b \in \mathbb{C}^{n}$, and  $A \in \mathbb{C}^{n \times n}$. 
Here, we assume $A$ is regular and a Hermitian matrix. 
%We keep to use a language invoking applied mathematics or linear computation, while detailed proofs for conjugate gradient (CG) and shifted CG solver are skipped.
%The rigorous proof can be found in \cite{SaadCG} and \cite{FrommerSBiCG}.

Next we introduce a key concept to govern non-stationary iterative solvers for linear equations, Krylov subspace.
Krylov subspace with $k$-th order is defined as,
\begin{align}
  \mathcal{K}_k(A,v)
  :=
  \mathrm{span} \left\{v,Av,\dots,A^{k-1}v\right\}.
\end{align}
Lower Krylov subspace is a subset of higher Krylov subspace:$\mathcal{K}_i \subset \mathcal{K}_j$ for $i<j$.

We suppose an iterative sequence to solve the linear equation starting from an initial guess $x_0$ and corresponding initial residual $r_0 = b-Ax_0$.
We seek a correction to initial guess by a rule such that the $k$-th correction is in $k$-th Krylov subspace with $v$ taken as the initial residual : $x_k - x_0 \in \mathcal{K}_k(A,r_0)$.
This procedure leads to a property that $k$-th residual is in the $k+1$-th Krylov subspace: $r_k \in \mathcal{K}_{k+1}(A,r_0)$.

The CG solver is derived such that $k$-th residual vector, $r_k = b-Ax_k$, is imposed to be perpendicular to the $k$-th subspace:
\begin{align}
  r_k
  \perp
  \mathcal{K}_k (A,r_0),
\end{align}
this is called Ritz-Galerkin condition.
%This condition determines a recursive formula for the residuals and hogehoge vectors.
This conditions leads to the well-known CG algorithm written in Table \ref{table:CG}.
\begin{table}[t]
  \caption{A pseudocode of CG algorithm, with a Hermitian matrix $A \in \mathbb{C}^{n\times n}$. $x_{k},r_{k},p_{k} \in \mathbb{C}^{n}$,  and $\alpha_{k},\beta_{k} \in \mathbb{C}$.
  }
  \label{table:CG}
  \begin{center}
    \begin{tabular}{ c l }
      \hline 
      1. & Set $x_{0}$, $p_{0}=r_{0}=b-Ax_{0}$ \\
      2. & For $k=0,1,\dots$ until convergence Do: \\
      3. & ~~~~~~$\alpha_k=(r_{k},r_{k})/(p_k,Ap_k)$ \\
      4. & ~~~~~~$x_{k+1}=x_{k}+\alpha_k p_k$ \\
      5. & ~~~~~~$r_{k+1} = r_k - \alpha_k Ap_k$ \\
      6. & ~~~~~~$\beta_k=(r_{k+1},r_{k+1})/(r_k,r_k)$ \\
      7. & ~~~~~~$p_{k+1} = r_{k+1} + \beta_k p_k$ \\
      8. & End Do \\
    \hline 
    \end{tabular}
  \end{center}
\end{table}
One should note that only linear operations to vector are included in the iteration, namely just update the vectors via scalar- or matrix-vector products.

Popular iterative solvers, the Biconjugate gradient (BiCG), the conjugate orthogonal conjugate gradient (COCG), the 
generalized minimal residual (GMRES), and so on, are also derived based on the concept of Krylov subspace with different matrices and different condition rather than Ritz-Galerkin condition.

%It seems, however, that solving linear equations to obtain $G_{\alpha \beta}(z)$ in Eq.~(\ref{eq:gnm}) has a critical problem, since 
%one has to solve the different linear equations when changing $z$. 
%To overcome this problem, the shifted COCG method has been proposed\cite{Takayama}.

\subsection{Shifted Conjugate Gradient (Shifted CG) method and shifted Krylov solvers}
We introduce a class of problem that series of linear equations whose matrices are connected each other by just shift of scalar times unit matrix from a {\it common} matrix with fixed right side vector:
\begin{align}
  \left(\sigma I + A\right) x(\sigma)
  =
  b,
\end{align}
with $\sigma \in \mathbb{C}$, a general regular matrix $A\in\mathbb{C}^{n\times n}$, and the unit matrix $I$.
This set of series is called shifted linear equations, or shifted linear systems in more mathematical community, and the scalar $\sigma$ is called shift.
One recognizes that this is frequently appeared in problem of physics, if the shift is regarded as a frequency.

Since Krylov subspace is the fixed vector multiplied by a polynomial composed by the matrix, Krylov subspace is invariant under the scalar shift with fixed vector:
\begin{align}
  \mathcal{K}_k (\sigma I + A,v)
  =
  \mathcal{K}_k (A,v).
\end{align}
This property allows us to reuse Krylov subspace created at a point to other shifted linear equation to obtain sequence of approximated solutions.
In the CG procedure, most computationally demanding parts are from the subspace construction, namely matrix-vector product, fifth line in the table \ref{table:CG}.
By sharing Krylov subspace constructed at {\it seed} shift, the most demanding parts can be skipped to construct the sequence of approximated solution at other shifts.
Solvers based on this idea are called shifted Krylov solvers.
We just focus on shifted CG solver for our purpose, while all of Krylov subspace solver are candidate to get the significant reduction of calculation cost.

\begin{widetext}
Hereafter, we proceed to constructing the shifted CG algorithm.
We assume Hermitian matrix for $A$ again and a real shift $\sigma \in \mathbb{R}$ here.
We enforce same initial residual to whole shifts, to realize same Krylov subspace construction.
A trivial solution is that zero vectors as initial guess for all shifts: $x_{0}(\sigma)=0, r_0(\sigma)=b$.
For simplicity, we take a point that $\sigma=0$ as the seed to generate Krylov subspace as an example.
Combining fifth and seventh lines in Table \ref{table:CG}, we obtain a recursive formula constituted by only residuals for the seed:
\begin{align}
  r_{k+1}
  =
  \left[\left(1+\frac{\beta_{k-1}}{\alpha_{k-1}}\alpha_k\right)I - \alpha_k A\right]r_k
  +
  \frac{\beta_{k-1}}{\alpha_{k-1}}\alpha_k r_{k-1}\label{eq:resi}.
\end{align}
Since $\sigma I + A$ is Hermitian, we can apply CG algorithm Table \ref{table:CG} to each shifts of shifted linear equations.
The relation \eqref{eq:resi} is also true for shifted linear equations:
\begin{align}
&  r_{k+1}(\sigma)
  =
  \left[\left(1+\frac{\beta_{k-1}(\sigma)}{\alpha_{k-1}(\sigma)}\alpha_k(\sigma)\right)I - \alpha_k(\sigma)\left(\sigma I+ A\right)\right]r_k(\sigma) 
  +
  \frac{\beta_{k-1}(\sigma)}{\alpha_{k-1}(\sigma)}\alpha_k(\sigma) r_{k-1}(\sigma)\label{eq:shiftedresi}.
\end{align}
Since generated Krylov subspace is common among whole shifts and dimension of a space such that $\mathcal{K}_{k+1} \cap \mathcal{K}_k$ is unity, the residuals at arbitrary shifts undergo collinear to the seed residual, $r_k(\sigma) \parallel r_{k}$.
We denote the coefficient to the seed residual as $r_k(\sigma) = \rho_k(\sigma) r_k$, which connects \eqref{eq:resi} and \eqref{eq:shiftedresi}.
Thus, we have formulae to obtain $\alpha_k(\sigma),\beta_k(\sigma)$ by comparing each components, $r_k, Ar_k, r_{k-1}$:
\begin{align}
&  \alpha_{k}(\sigma)=\frac{\rho_{k+1}(\sigma)}{\rho_{k}(\sigma)} \alpha_{k},\quad
  \beta_{k}(\sigma)=\left( \frac{\rho_{k+1}(\sigma)}{\rho_{k}(\sigma)} \right)^{2} \beta_{k},\quad 
  \rho_{k+1}(\sigma) =\frac{\rho_{k}(\sigma)\rho_{k-1}(\sigma) \alpha_{k-1}}{\rho_{k-1}(\sigma) \alpha_{k-1}(1+\alpha_{k} \sigma)+\alpha_{k} \beta_{k-1}(\rho_{k-1}(\sigma)-\rho_{k}(\sigma))}.
\end{align}
This relation allows us to evolve the iterations at shifts without any matrix-vector products via rational operation for scalar quantities obtained at the seed point.
This is shifted CG.
\end{widetext}

The shifted CG is not applicable to $G_{\alpha \beta}(z)$ in Eq.~(\ref{eq:gnm}), since complex value $z$ violates the Hermitian condition for the matrix $zI-H$.
For real symmetric common matrix with complex shift, shifted COCG is suitable\cite{Takayama}.
However, shifted CG algorithm can be applied to a problem that shifts are generally complex as long as we take the seed on real-axis.
We explain how the trick is justified in our problem.

We assume more general situation, a regular matrix $A$ and complex shift $\sigma$, for a while.
For this purpose, We have shifted biconjugate gradient (BiCG) solver which is applied to linear response calculation within time-dependent density-functional theory\cite{Hueberner}.
%A key structure in CG, COCG and BiCG is that the recursive formulae composes only linear operations for vector quantities.
The shifted BiCG solver requires an additional auxiliary problem $\left(\sigma I + A\right)^H y(\sigma) = c$, where choice of the vector $c$ is arbitrary unless orthogonal to the vector in original problem $c^H b = 0$.
The shifted BiCG algorithm is quite similar to the shifted CG one except for evaluation ways to $\alpha_k$ and $\beta_k$ at seed.
When we take a condition that the matrix $A$ is Hermitian and shift is real and $c=b$, the seed algorithm of shifted BiCG becomes exactly same as one of shifted CG.
Additionally, generation formulae for shifts from a seed is common among the two approaches.
Therefore, we can take a approach that CG iteration is evolved at a seed on real axis and approximated solutions at generally complex shifts are generated according to effectively same formula as shifted CG.
The actual procedure is written in Table \ref{table:SCG}.
\begin{table*}[hbt]
  \caption{Shifted CG for $(\sigma_{j} I + A) x(\sigma_{j}) = b$ with a hermitian matrix $A \in \mathbb{C}^{n \times n}$ and complex shift $\sigma_j \in\mathbb{C}$. $x_{k},r_{k},p_{k} \in \mathbb{C}^{n}$, $\alpha_{k}(\sigma_{j}),\beta_{k}(\sigma_{j}),\rho_{k}(\sigma_{j})\in\mathbb{C}$, $x_{k}(\sigma_{j}),p_{k}(\sigma_{j})\in\mathbb{C}^n$.
  }
  \label{table:SCG}
  \begin{center}
    \begin{tabular}{ c l }
      \hline 
      1. & Input $\sigma_{j}$($j = 1,2,\cdots,N$) \\
      2. & Set $x_{0}=0$, $r_{0}=p_{0}=b$, $\alpha_{-1}=1$,$\beta_{-1}=0$ \\
      3. & Set $x_{0}(\sigma_{j})=0$, $r_{0}(\sigma_{j})=p_{0}(\sigma_{j})=b$, $\rho_{-1}(\sigma_{j})=\rho_{0}(\sigma_{j})=1$ ($j=1,2,\cdots,N$) \\
      4.  & For $k=0,1,\dots$ until convergence Do: \\
      5.  & ~~~~~~$\alpha_k=(r_{k},r_{k})/(p_k,Ap_k)$ \\
      6.  & ~~~~~~$x_{k+1}=x_{k}+\alpha_k p_k$ \\
      7.  & ~~~~~~$r_{k+1} = r_k - \alpha_k Ap_k$ \\
      8.  & ~~~~~~$\beta_k=(r_{k+1},r_{k+1})/(r_k,r_k)$ \\
      9. & ~~~~~~$p_{k+1} = r_{k+1} + \beta_k p_k$ \\
      10.  & ~~~~~~For $j = 1,2,\cdots,N$ Do: \\
      11.  & ~~~~~~~~~~~~$\rho_{k+1}(\sigma_{j}) =\frac{\rho_{k}(\sigma_{j})\rho_{k-1}(\sigma_{j})\alpha_{k-1}}{\rho_{k-1}(\sigma_{j}) \alpha_{k-1}(1+\alpha_{k} \sigma_{j})+\alpha_{k} \beta_{k-1} (\rho_{k-1}(\sigma_{j})-\rho_{k}(\sigma_{j})) }$\\
      12.  & ~~~~~~~~~~~~$\alpha_{k}(\sigma_{j})=\frac{\rho_{k+1}(\sigma_{j})}{\rho_{k}(\sigma_{j})} \alpha_{k}$ \\
      13.  & ~~~~~~~~~~~~$x_{k+1}(\sigma_{j})=x_{k}(\sigma_{j}) +\alpha_{k}(\sigma_{j}) p_{k}(\sigma_{j})$ \\
      14.  & ~~~~~~~~~~~~$\beta_{k}(\sigma_{j})=\left( \frac{\rho_{k+1}(\sigma_{j})}{\rho_{k}(\sigma_{j})} \right)^{2} \beta_{k}$ \\
      15.  & ~~~~~~~~~~~~$p_{k+1}(\sigma_{j})=r_{k+1}(\sigma_{j})+\beta_{k}(\sigma_{j}) p_{k}(\sigma_{j})$ \\
      16.  & ~~~~~~~~~~~~$r_{k+1}(\sigma_{j})=\rho_{k+1}(\sigma_{j}) r_{k+1}$ \\
      17. & ~~~~~~End Do \\
      18. & End Do \\
    \hline 
    \end{tabular}
  \end{center}
\end{table*}

This trick allows us to reduce computational effort in two situations.
First situation is that complex shifts with Hermitian common matrix, $A$.
The operations related to auxiliary linear equation in BiCG can be skipped, leading to half reduction of most computationally demanding  part, namely matrix-vector operation at seed.
Second situation is that complex shifts with real symmetric matrix for common matrix rather than Hermitian.
For the case, we have another choice that employing COCG and shifted COCG, regardless the trick.
However, COCG requires complex component operations if we take a seed on not real-axis.
Only real-value operations are available by putting a seed on real axis, leading to half reduction of the matrix-vector operation.

%A shifted BiCG, a cousin of shifted CG, is success for optical response within time-dependent density-functional theory \cite{Hueberner}
%A key structure in CG, COCG and BiCG is that the recursive formulae composes only linear operations for vector quantities.
Although significant reduction of heaviest calculation cost is realized by use of the shifted CG, it does not help to reduce memory consumption at all.
Since many shift points, physically Matsubara frequencies, are required for our problem, the large amount of memory consumption secondly limits us to obtain Green's function \eqref{eq:gnm} for large scale simulation.

%We should note, however, that the shifted (CO)CG method also has a critical problem to calculate the Green's function in Eq.~(\ref{eq:gnm}); the computational memories required are proportional to 
%the product of the dimension of the vector $|m \rangle$ by the number of different $z$ that one needs. 
%This problem restricts us from calculating large number of the elements with different frequencies $z$ when treating a large Hilbert space. 
\subsection{Reduced-Shifted Conjugate Gradient (RSCG) method}
To overcome the above difficulties to calculate a Green's function with linear equations, 
we propose an efficient approach to obtain reduced vector quantities for shifts evaluated by a fixed matrix times solution vectors, we call it reduced-shifted conjugate gradient (RSCG).
The reduced vector quantities are defined as 
\begin{align}
  \Xi(\sigma) = Vx(\sigma)
\end{align}
with $\Xi(\sigma)\in \mathbb{C}^m$, $V\in \mathbb{C}^{m\times n}$, and $x(\sigma)\in \mathbb{C}^n$, derived from shifted linear equations $\left(\sigma I + A\right)x(\sigma)=b$.

It is frequently appeared that solutions are not needed but the only reduced quantities are needed as physical quantities, {\it e.g.} a reduced value from a Green's function $\int dx f^*(x)\left[\left(\omega - \hat{h}\right)^{-1}g\right](x)$.
Matrix elements of the Green's function with different indices $j$ and a fixed index $i$ in Eq.~(\ref{eq:gnm}) is calculated by these reduced vector quantities. 
Since shifted CG include linear operation to evolve the iteration, we can construct an algorithm such that evolving recurrence formulae by only use of reduced quantities defined as,
\begin{align}
  \Xi_k(\sigma)=V x_k(\sigma),\quad
  \Pi_k(\sigma)=V p_k(\sigma),\quad
  \Sigma_k = V r_k,
\end{align}
with $\Pi_k(\sigma),\Sigma_k \in \mathbb{C}^m$.
The reduced quantities are not evaluated explicitly at each iteration but at just preparation stage only once before iteration except for $\Sigma$, namely $\Xi_0(\sigma)=V 0=0, \Pi_0(\sigma)=V p_0(\sigma)=V b$.
Then, the iteration proceeds according to alternative recurrences,
\begin{align}
&  \Sigma_k
  =
  V r_k,\qquad
  \Xi_{k+1}(\sigma)
  =
  \Xi_{k}(\sigma) + \alpha_k(\sigma)\Pi_k,\qquad \nonumber \\
&  \Pi_{k+1}(\sigma)
  =
  \rho(\sigma)\Sigma_k + \beta_k(\sigma)\Pi_k.
\end{align}
corresponding to thirteenth, fifteenth, and sixteenth line in Table \ref{table:SCG}, which is clearly written as closed form among reduced quantities.
A pseudocode of the RSCG is written in Table \ref{table:RSCG}.
\begin{table*}[hbt]
  \caption{Reduced-shifted CG for $(\sigma_{j} I + A) x(\sigma_{j}) = b$ with a hermitian matrix $A \in \mathbb{C}^{n \times n}$. $V \in \mathbb{C}^{m \times n}$, $x_{k},r_{k},p_{k} \in \mathbb{C}^{n}$, 
  $\alpha_{k}(\sigma_{j}),\beta_{k}(\sigma_{j}),\rho_{k}(\sigma_{j}),\Xi_{k}(\sigma_{j}),\Pi_{k}(\sigma_{j}),\Sigma_{k}  \in \mathbb{C}^{m}$, and $\alpha_{k},\beta_{k} \in \mathbb{C}$.
  $\Xi_{k}(\sigma_{j}) \equiv V x_{k}(\sigma_{j})$. $V^{\rm T} \equiv (v_{1},v_{2},\cdots,v_{m})$. $v_{i} \in \mathbb{C}^{n}$. 
  Here, the symbol T represents transposition. 
  }
  \label{table:RSCG}
  \begin{center}
    \begin{tabular}{ c l }
      \hline 
      1.  & Input $\sigma_{j}$($j = 1,2,\cdots,N$) \\
      2. & Set $x_{0}=0$, $r_{0}=p_{0}=b$, $\alpha_{-1}=1$,$\beta_{-1}=0$ \\
      3. & Compute $\Sigma_{0} = V b$ \\
      4. & Set $\Xi_{0}(\sigma_{j})=0$, $\Pi_{0}(\sigma_{j})=\Sigma_{0}$, $\rho_{-1}(\sigma_{j})=\rho_{0}(\sigma_{j})=1$ ($j=1,2,\cdots,N$) \\
      5.  & For $k=0,1,\dots$ until convergence Do: \\
      6.  & ~~~~~~$\alpha_k=(r_{k},r_{k})/(p_k,Ap_k)$ \\
      7.  & ~~~~~~$x_{k+1}=x_{k}+\alpha_k p_k$ \\
      8.  & ~~~~~~$r_{k+1} = r_k - \alpha_k Ap_k$ \\
      9.  & ~~~~~~$\beta_k=(r_{k+1},r_{k+1})/(r_k,r_k)$ \\
      10. & ~~~~~~$p_{k+1} = r_{k+1} + \beta_k p_k$ \\
      11. & ~~~~~~Compute $\Sigma_{k+1} = V r_{k+1}$ \\
      12.  & ~~~~~~For $j = 1,2,\cdots,N$ Do: \\
      13.  & ~~~~~~~~~~~~$\rho_{k+1}(\sigma_{j}) =\frac{\rho_{k}(\sigma_{j})\rho_{k-1}(\sigma_{j}) \alpha_{k-1}}{\rho_{k-1}(\sigma_{j}) \alpha_{k-1}(1+\alpha_{k} \sigma_{j})+\alpha_{k} \beta_{k-1} (\rho_{k-1}(\sigma_{j})-\rho_{k}(\sigma_{j})) }$\\
      14.  & ~~~~~~~~~~~~$\alpha_{k}(\sigma_{j})=\frac{\rho_{k+1}(\sigma_{j})}{\rho_{k}(\sigma_{j})} \alpha_{k}$ \\
      15.  & ~~~~~~~~~~~~$\Xi_{k+1}(\sigma_{j})=\Xi_{k}(\sigma_{j}) +\alpha_{k}(\sigma_{j}) \Pi_{k}(\sigma_{j})$ \\
      16.  & ~~~~~~~~~~~~$\beta_{k}(\sigma_{j})=\left( \frac{\rho_{k+1}(\sigma_{j})}{\rho_{k}(\sigma_{j})} \right)^{2} \beta_{k}$ \\
      17.  & ~~~~~~~~~~~~$\Pi_{k+1}(\sigma_{j})=\rho_{k+1}(\sigma_{j})\Sigma_{k+1}+\beta_{k}(\sigma_{j}) \Pi_{k}(\sigma_{j})$ \\
      18. & ~~~~~~End Do \\
      19. & End Do \\
    \hline 
    \end{tabular}
  \end{center}
\end{table*}
The matrix $V$ is chosen as expected physical quantities $v_i^T x(\sigma)$ corresponding to $v_i$ with $V^T=(v_1,v_2,\dots,v_m)$.

The RSCG is favorable due to significant reduction of memory consumption, because typically number of dimension of reduced quantities is significantly smaller than dimension of linear problem: $m\ll n$.
While usual shifted CG requires that three vectors on each shifts is stored in memory, RSCG does just three reduced vectors.
The reduction of relevant quantities helps us reduction of the calculation cost as well.
When number of shifts are close to problem dimension $n$, a cost to evolve iteration for whole shifts, large amount of scalar-vector products,  becomes comparable to a cost to construct a Krylov subspace on a seed.
By use of RSCG, the calculation is reduced to scalar times reduced vector, $m$, from one times full-dimension vector, $n$.
While the RSCG seems almost trivial to work properly from algebraic point of view as long as shifted CG is working, there has been no application to actual problems.
%While the RSCG seems almost trivial to work from algebraic point of view, there has been no application to actual problems.

Finally, we describe how to calculate the Green's function with the use of the RSCG method.
We consider $m$ complex frequencies $z_{j}$. 
The shift points are defined as $\sigma_{j} = z_{j}$. 
We set $A = - H$. 
The matrix element of the Green's function $G_{\alpha \beta}(z_{j})$ is defined in Eq.~(\ref{eq:gnm}).  
In the RSCG method, the matrix elements with different indices $n$ and a fixed $m$ are simultaneously solved. 
Thus, we define the $m \times n$ matrix $V$ defined as 
\begin{align}
V^{\rm T} &= \left(\begin{array}{cccc}\langle 1| & \langle 2| & \cdots & \langle m| \end{array}\right).
\end{align}
Here, the symbol T represents transposition and $\langle \alpha |$ is a $n$ dimensional vector. 
The matrix elements $G_{1\beta}^{k}(z_{j}),G_{2\beta}^{k}(z_{j}),\cdots,G_{m\beta}^{k}(z_{j})$ on the $k$-th iteration step are calculated as elements of $m$-dimensional vector $\Xi_{k}(z_{j})$
\begin{align}
\Xi_{k}(z_{j}) &= 
\left(\begin{array}{cccc}
G_{1\beta}^{k}(z_{j})
 & G_{2\beta}^{k}(z_{j}) &
  \cdots & G_{m\beta}^{k}(z_{j}) \end{array}\right), 
\end{align}
by putting 
$b = |\beta \rangle$ in Table \ref{table:RSCG}. 
The RSCG iteration loop is stopped when the residual on the $k$-step  
\begin{align}
{\rm res}_{k} &= {\rm Max} ||r_{k}(z_{j})||,
\end{align}
with 
\begin{align}
r_{k}(z_{j}) &= \rho_{k}(z_{j}) r_{k}, \label{eq:rkr}
\end{align}
reaches a desired accuracy.

%we propose the RSCG method, by focusing on a matrix element of the Green's function. 

\section{Formulation for the Bogoliubov-de Gennes theory of superconductivity}\label{sec:bdg}
We focus on the formulation for the Bogoliubov-de Gennes theory of superconductivity. 
We introduce the self-consistent calculation scheme with the use of the RSCG method. 
\subsection{Hamiltonian and gap equation}
Throughout this paper, we set $\hbar = k_{\rm B} = 1$. 
Let us consider a meanfield Bardeen-Cooper-Schrieffer (BCS) Hamiltonian:
%We consider a weak-coupling BdG Hamiltonian:
\begin{align}
 H 
= 
\frac{1}{2}\psi^{\dagger}\hat{H}\psi
=
\frac{1}{2}
(\bar{c}^{\rm T},c^{\rm T})
\left(
\begin{array}{cc}
\hat{H}^{\rm N}         &  \hat{\Delta} \\
\hat{\Delta}^{\dagger} & -\hat{H}^{{\rm N} \ast}
\end{array}
\right)
\left(
\begin{array}{c}
c \\
\bar{c}
\end{array}
\right),
\end{align}
with 
\mbox{
\(
c=(c_{1},c_{2},\ldots,c_{N})^{\rm T}
\)} 
and 
\mbox{
\(
\bar{c}=(c_{1}^{\dagger},c_{2}^{\dagger},\ldots,c_{N}^{\dagger})^{\rm T}
\)}. 
The fermionic annihilation and creation operators are denoted as,
respectively, $c_{i}$ and $c_{i}^{\dagger}$ ($i=1,\ldots,N$). 
The index $i$ includes all the relevant degrees of freedom such as
spatial sites, spins, orbitals, and so on. 
The canonical anti-commutation relation is 
\mbox{$[c_{i},c_{j}^{\dagger}]_{+}=\delta_{ij}$}. 
The Hamiltonian matrix $\hat{H}$ is a $2 N \times 2 N$ Hermitian matrix. 
$\hat{\Delta}$ corresponds to the superconducting order parameter. 
The superconducting gap equation is given as 
\begin{align}
\Delta_{ij} &= \sum_{k l} U_{ijkl} \langle c_{k} c_{l} \rangle, 
\end{align}
where $\langle c_{k} c_{l} \rangle$ is a superconducting Cooper pair meanfield defined below.  
Here, $U_{ijkl}$ is a general pairing interaction. 
For example, in the single-band $s$-wave and $d$-wave superconductors, the pairing interaction is simplified as $U_{ijkl} = \delta_{ik} \delta_{jl} U_{ij}$.

\subsection{Bogoliubov-de Gennes equations}
%The BdG equations are regarded as the eigenvalue equations with respect to $\hat{H}$
%
The Hamiltonian $H$ is diagonalized by solving the corresponding BdG equations written as 
\begin{align}
\sum_{j} 
\left(
\begin{array}{cc}
\left[ \hat{H}^{\rm N} \right]_{ij}         &  \left[\hat{\Delta} \right]_{ij} \\
\left[ \hat{\Delta}^{\dagger} \right]_{ij}&
 -\left[ \hat{H}^{{\rm N} \ast} \right]_{ij}
\end{array}
\right)
\left(
\begin{array}{c}
u^{\alpha}_{j} \\
v^{\alpha}_{j}
\end{array}
\right)
&= E^{\alpha} 
\left(
\begin{array}{c}
u^{\alpha}_{i} \\
v^{\alpha}_{i}
\end{array}
\right).
\end{align}
To solve the BdG equations is equivalent to the diagonalization of $\hat{H}$ with a unitary matrix $\hat{U}$.
The matrix elements of $\hat{U}$ are 
\begin{align}
U_{i,\alpha} &= u^{\alpha}_{i}, \\
U_{i+N,\alpha} &= v^{\alpha}_{i}.
\end{align}
The eigenvalues $E^{\alpha}$ are not independent of each other. 
With the use of the particle-hole transformation, $-E^{\alpha}$ is also an eigenvalue and its eigenvector is $(\Vec{v}^{\alpha \ast}, \Vec{u}^{\alpha \ast})^{\rm T}$. 

\subsection{Eigenvalue distributions and Kernel polynomial method}
In order to consider a nano-fabricated superconductor, 
we introduce  the ``fabrication'' potential $\hat{V}$ whose elements are quite large to carve the shape of a nano structure.  
Thus, $\hat{H}^{\rm N}$ is expressed as 
$\hat{H}^{\rm N} = \hat{H}^{\rm N,0} + \hat{V}$,
with the normal-state Hamiltonian $\hat{H}^{\rm N,0}$ and $[\hat{V}]_{ij} = V_{i} \delta_{ij}$. 
We show that the Hamiltonian matrix with the fabrication potential $\hat{V}$ has a wide-range of the eigenvalue distribution. 
According to the Gershgorin's circle theorem\cite{Golub}, every eigenvalue of $n \times n$ matrix $A$ lies within at least one of the Gershgorin discs $D(a_{ii},R_{i})$. 
Here, $D(a_{ii},R_{i})$ is the closed disc centered at $a_{ii}$ with radius $R_{i}$. $a_{ij}$ denotes an element of the matrix $A$ and $R_{i} = \sum_{j \neq i} |a_{ij}|$ is the 
sum of the absolute values of the non-diagonal element in the $i$-th row. 
Since the absolute value of the diagonal element becomes large by introducing $\hat{V}$, the eigenvalues are distributed in the range 
\begin{align}
- |{\rm Max}\:(V_{i})| \lesssim E^{\alpha}  \lesssim  |{\rm Max}\:(V_{i})|. 
\end{align}

This wide eigenvalue distribution prevents an accurate calculation with the use of the kernel polynomial method. 
In the KPM, the delta function is expanded by orthonormal polynomials in [-1,1]. 
Thus, one has to rescale $\hat{H}$ and $\omega$ so that $\hat{K} = (\hat{H} -b)/a$ and $x=(\omega - b)/a$, with $a = (E_{\rm max} - E_{\rm min})/2$ and 
$b = (E_{\rm max} + E_{\rm min})/2$. 
The energy resolution of this method is defined by $a/n_{\rm K}$ with the polynomial cutoff parameter $n_{\rm K}$.  
With increasing the fabrication potential, the energy resolution decreases with a fixed $n_{\rm K}$. 
Therefore, the KPM is not suitable for a nano-fabricated superconductor.

\subsection{Green-function-based self-consistent calculation}
Mean fields are calculated by one-particle Green's functions. 
The $2 N \times 2 N$ matrix Green's function with a complex frequency is defined as 
\begin{align}
\hat{G}(z) &= \left[z\hat{I} - \hat{H} \right]^{-1}.
\end{align}
With the unitary matrix $\hat{U}$, each component of $\hat{G}(z)$ is expressed as 
\begin{align}
G_{\alpha \beta}(z) &= \sum_{\gamma=1}^{2N} U_{\alpha \gamma} U_{\beta \gamma}^{\ast} \frac{1}{z - E^{\gamma}}, \: (1 \le \alpha,\beta \le 2N) \label{eq:direc}
\end{align}
If we set $z = i \omega_{n}$ with the Matsubara frequency $\omega_{n} = (2 n +1) \pi T$, the above formula corresponds to the Matsubara temperature Green's function. 
The retarded and advanced Green's functions are, respectively, defined as 
\begin{align}
\hat{G}^{\rm R}(\omega) &= \lim_{\eta \rightarrow 0+} \hat{G}(\omega + i \eta),\\
\hat{G}^{\rm A}(\omega) &= \lim_{\eta \rightarrow 0+} \hat{G}(\omega - i \eta).
\end{align}
In order to obtain physical quantities (e.g. density of states) from Green's functions, we introduce the following useful $2N$-component unit-vectors
$\Vec{e}(i)$ and $\Vec{h}(i)$ ($i \le i \le N$), which are, respectively, defined as 
\begin{align}
[\Vec{e}(i)]_{\gamma} &= \delta_{i,\gamma}, \: [\Vec{h}(i)]_{\gamma} = \delta_{i+N,\gamma}. 
\end{align}
For example, the local density of states with respect to the site $i$ is given as
\begin{align}
N(\omega,i) &= - \frac{1}{2 \pi i}  \Vec{e}(i)^{\rm T} \hat{d}(\omega) \Vec{e}(i), 
\end{align}
with use of the spectral function $\hat{d}(\omega)$ defined as 
\begin{align}
\left[ \hat{d}(\omega) \right]_{\alpha \beta} &\equiv \left[ \hat{G}^{\rm R}(\omega) -\hat{G}^{\rm A}(\omega) \right]_{\alpha \beta} \\
&= - 2 \pi i \sum_{\gamma=1}^{2N} U_{\alpha \gamma} U^{\ast}_{\beta \gamma} \delta(\omega - E^{\alpha}). \label{eq:du}
\end{align}
Two types of mean fields $\langle c_{i}^{\dagger} c_{j} \rangle$ and $\langle c_{i} c_{j} \rangle$ can be expressed as 
\begin{align}
\langle c_{i}^{\dagger} c_{j} \rangle &= - \frac{1}{2 \pi i} \int^{\infty}_{-\infty} d \omega f(\omega) \Vec{e}(j)^{\rm T} \hat{d}(\omega) \Vec{e}(i), \\
\langle c_{i} c_{j} \rangle &= - \frac{1}{2 \pi i} \int^{\infty}_{-\infty} d \omega f(\omega) \Vec{e}(j)^{\rm T} \hat{d}(\omega) \Vec{h}(i),
\end{align}
with $f(x) \equiv 1/(e^{x/T} +1)$. 
With the use the analytic continuation, these mean fields are rewritten as 
\begin{align}
\langle c_{i}^{\dagger} c_{j} \rangle &= T \sum_{n=-\infty}^{\infty}  \Vec{e}(j)^{\rm T} \hat{G}(i \omega_{n}) \Vec{e}(i), \\
\langle c_{i} c_{j} \rangle &= T \sum_{n=-\infty}^{\infty}  \Vec{e}(j)^{\rm T} \hat{G}(i \omega_{n}) \Vec{h}(i). 
\end{align}
By solving the linear equations defined as 
\begin{align}
(i \omega_{n} \hat{I} -\hat{H}) \Vec{x}(i,\omega_{n}) &= \Vec{h}(i), \label{eq:meanx}
\end{align}
the superconducting mean field is expressed as 
\begin{align}
\langle c_{i} c_{j}\rangle &= T \sum_{n=-n_{\rm c}}^{n_{c}} \Vec{e}(j)^{\rm T} \Vec{x}(i,\omega_{n}).
\end{align}
Here, $n_{\rm c}$ is the Matsubara cutoff parameter. 
The residual of the RSCG method of $k$-th iteration is defined as 
\begin{align}
{\rm res}_{k} &= {\rm Max} || \Vec{r}_{k}(\omega_{n}) ||, 
\end{align}
with
\begin{align}
 \Vec{r}_{k}(\omega_{n}) &= \Vec{h}(i)- (i \omega_{n} \hat{I} -\hat{H}) \Vec{x}_{k}(i,\omega_{n}). \label{eq:rk}
\end{align}
The residual vector $\Vec{r}_{k}(\omega_{n})$ is estimated by $\rho_{k}(\omega_{n}) \Vec{r}_{k}$. 
Here, $\Vec{r}_{k} = \Vec{h}(i)+ \hat{H} \Vec{x}_{k}(i, 0)$ and the value $\rho_{k}(\omega_{n})$ is calculated in the each iteration step.

Thus, the self-consistent  cycle has the schematic form 
\begin{align}
\hat{\Delta} &\rightarrow \hat{G}(i \omega_{n}) \nonumber \\
&\rightarrow \langle c_{i}^{\dagger} c_{j} \rangle \\
&\rightarrow \hat{\Delta}. \nonumber
\end{align}

\subsection{Calculation of physical quantities}
Once one obtains a converged order parameter $\hat{\Delta}$, 
there are two choices to calculate physical quantities, the RSCG method on a real-frequency axis or the Sakurai-Sugiura(SS) method\cite{Sakurai}.
The RSCG method can be used for calculations of various kinds of physical quantities defined on a real-frequency axis. 
For example, the local density of states $N(\omega,i)$ is obtained
\begin{align}
N(\omega,i) &= - \frac{1}{\pi} {\rm Im} \: \left[ 
 \Vec{e}(i)^{\rm T} \Vec{x}(i,\omega+i \eta)
\right],
\end{align}
by solving the following linear equation
\begin{align}
\left[\omega + i \eta - \hat{H} \right]  \Vec{x}(i,\omega+i \eta) &= \Vec{e}(i). 
\end{align}
with a smearing factor $\eta$.

The Sakurai-Sugiura (SS) method is also appropriate when the eigenvalues and eigenfunctions of the BdG equations are needed\cite{Sakurai}. 
This method is a numerical solver for a generalized eigenvalue problem so that $A \Vec{x} = \epsilon B \Vec{x}$ and has been applied to 
various physical issues such as 
the real-space density functional theory\cite{Futamura}, the lattice quantum chromodynamics\cite{Ohno}, and inhomogeneous superconductors\cite{NagaiSS}.
The SS method allows us to extract the eigen-pairs whose eigenvalues are located in a given domain on the complex plane, from a generic matrix\cite{zPares}.

\section{Numerical demonstrations for superconductivity} \label{sec:demo}
To illustrate power of our method, we show the results of $d$-wave nano-island and $s$-wave vortex lattice as examples.
In this section, we show that the RSCG method can treat the realistic length scales  by considering relatively large size systems. 
As we noted, superconductivity has multi length scales (i.e. an inverse of a Fermi-wave length $1/k_{\rm F}$ ($\sim$ \AA), a superconducting coherence length $\xi$ ($\sim$ 10-100 nm), and a magnetic field penetration length $\lambda$ ($\sim \mu$m)). 
Considering two-dimensional systems, we assume that the field penetration length $\lambda$ is infinity. 
In the quasiclassical Eilenberger theory, which is usually used for a large inhomogeneous superconductor, the oscillations characterized by Fermi-wave length $1/k_{\rm F}$ are neglected. 
In the conventional BdG calculations, the coherence length $\xi$ should be short $\xi \sim 1/k_{\rm F}$, since the coherence length should be shorter than the system size limited by the computational cost.

In both cases, we consider the single orbital square-lattice tight-binding model with a nearest neighbor hopping $t_{ij}$ expressed as 
\begin{align}
\left[ \hat{H}^{\rm N} \right]_{ij} &= -t_{ij} - \mu \delta_{ij}+ V_{i} \delta_{ij}.
\end{align}
We use the Matsubara cutoff parameter $n_{\rm c} = 23998$ such that $\omega_{\rm c} = 240 \pi = \pi T (2 n_{\rm c}+1)$, where the number of the total shifted points is $23998\times 2+1=47997$. 

\subsection{$d$-wave nano island}
We consider a $d$-wave nano island. 
The pairing interaction is $U_{ijkl} = U_{ij} \delta_{ik} \delta_{jl}$, where $U_{ij}$ denotes the pairing interaction given only on the link of the nearest neighbor and its amplitude is $U$. 
We self-consistently calculate $d_{x^{2}-y^{2}}$-wave order parameter 
\begin{align}
\Delta_{d,i} = U (\Delta_{\hat{x},i}+\Delta_{-\hat{x},i} - \Delta_{\hat{y},i} - \Delta_{-\hat{y},i})/4, \label{eq:deld}
\end{align}
with $\Delta_{\pm \hat{e}} = \Delta(\Vec{r}_{i},\Vec{r}_{i} \pm \Vec{e})$,
where 
$\Vec{r}_{i}$ is a position vector with index $i$ whose origin is a center of the system $\Vec{r}_{i} = (i_{x}-L_{x}/2-1/2,i_{y}-L_{y}/2-1/2)$ 
and $\hat{x}$ and $\hat{y}$ denote the unit vectors in a square lattice.
We set $[\hat{V}]_{ij} = \delta_{ij} V_{i}$ and $V_{i} = 100t$ on the site $i$ outside the circle located at $\Vec{r} = (0,0)$ with a radius of $3L_{x}/8$. 
We consider the chemical potential $\mu = -1.5t$, the pairing interaction $U = -2t$, the temperature $T = 0.01t$ and the system size $L_{x}\times L_{y} = 192 \times 192$, whose matrix dimension is $73728$. 
A criteria for the residual Eq. (\ref{eq:rkr}) is set to 0.1 in each linear equation.
%The convergence of the RSCG method is set to $0.1$.  

\begin{figure}[t]
%%%%--- I comment out figure regions
%\vspace{50mm}
\begin{center}
     \begin{tabular}{p{   \columnwidth}} %p{0.5 \columnwidth}}%  p{28mm}}
%          \begin{tabular}{p{  0.5 \columnwidth}p{ 0.5  \columnwidth}} %p{0.5 \columnwidth}}%  p{28mm}}
      (a)\resizebox{0.8 \columnwidth}{!}{\includegraphics{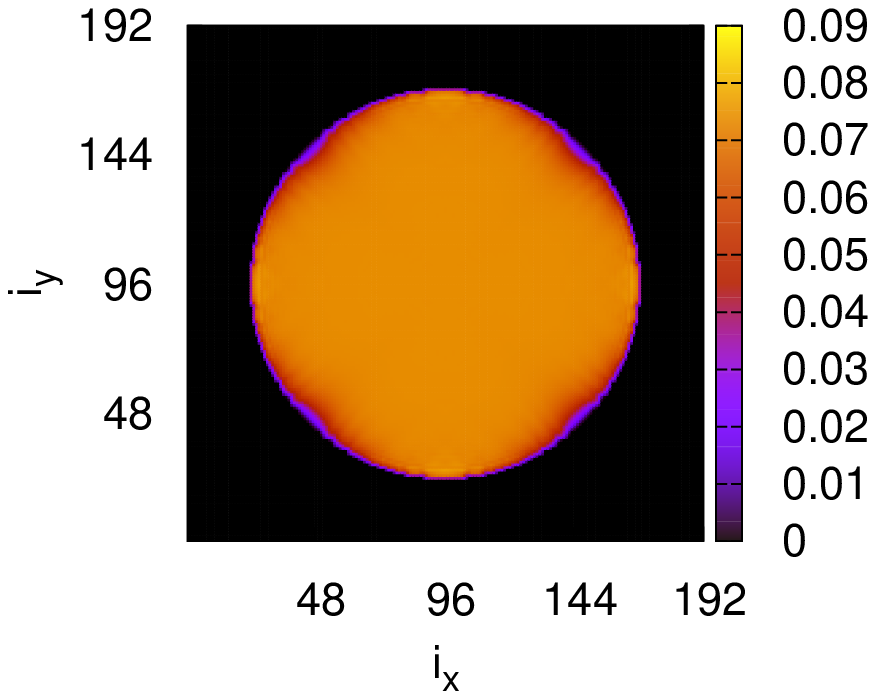}} \\
%      &
      %\\ %&
      (b)\resizebox{0.8 \columnwidth}{!}{\includegraphics{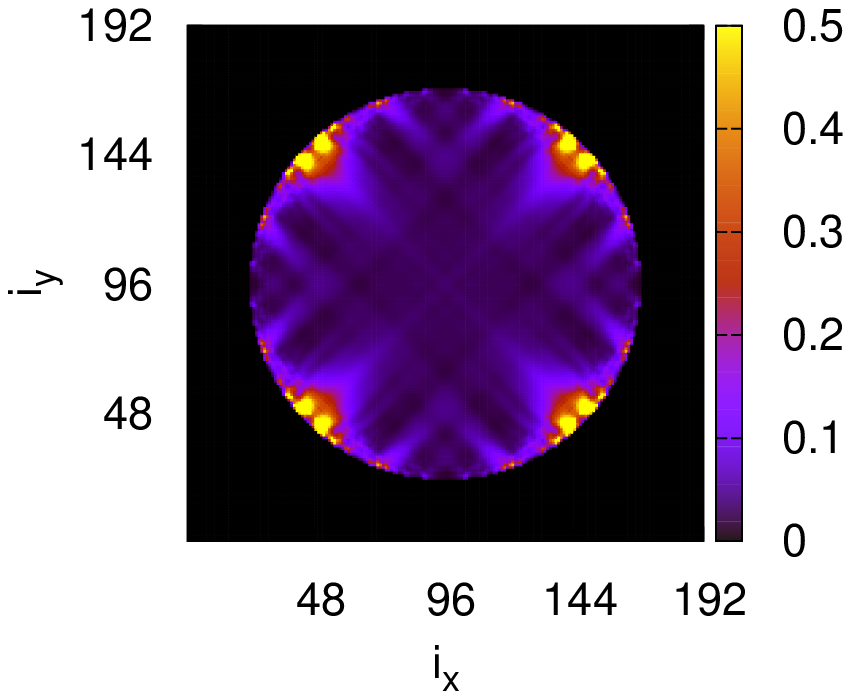}} 
    \end{tabular}
\end{center}
\caption{(Color online) 
Amplitude of the $d$-wave order parameter in the circler nano island. 
The chemical potential $\mu = -1.5t$ and the pairing interaction $U = -2t$. The system size is $L_{x} \times L_{y} = 192 \times 192$. We consider the temperature $T = 0.01t$ and the smearing factor $\eta = 0.01t$.
\label{fig:fig1}
 }
\end{figure}

Figure \ref{fig:fig1}(a) shows the converged order parameter defined in Eq.~(\ref{eq:deld}).
The order parameter is small at the 110 boundaries, where the Andreev bound states appear as shown in Fig.~\ref{fig:fig1}(b). 
Note that we use the SS method to calculate the zero energy density of states with the use of Eq.~(\ref{eq:du}). 
Here, we adopt the smearing factor $\eta = 0.01t$. 
These bound states are similar to the results in the system calculated by the quasiclassical Eilenberger theory\cite{NagaiNano}. 
In this BdG calculation, however, one can clearly see the quantum oscillations characterized by the inverse of the Fermi-wave length $1/k_{\rm F}$. 
Thus, we show that the RSCG method can treat two characteristic length scales $\xi$ and $1/k_{\rm F}$. 

\begin{figure}[t]
%%%%--- I comment out figure regions
%\vspace{50mm}
\begin{center}
     \begin{tabular}{p{ 0.8 \columnwidth}} %p{0.5 \columnwidth}}%  p{28mm}}
      \resizebox{0.8 \columnwidth}{!}{\includegraphics{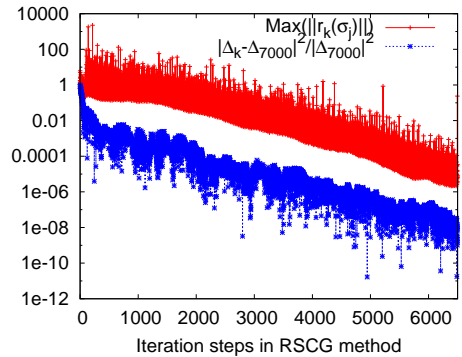}} 
%      &
      %\\ %&
 %     \resizebox{0.8 \columnwidth}{!}{\includegraphics{Fig1b.eps}} 
    \end{tabular}
\end{center}
\caption{
(Color online) 
The maximum of the residual-vector amplitudes ${\rm Max}\: ||r_{k}(\sigma_{j})||$ and the convergence profile of the $d$-wave order parameter at the center  $(i_{x},i_{y}) = (L_{x}/2,L_{y}/2)$. The parameters are same in Fig.~\ref{fig:fig1}. $\Delta_{k}$ denotes the $d$-wave order parameter at the center on the $k$-th iteration step in the RSCG method. 
\label{fig:fig2}
 }
\end{figure}

Figure \ref{fig:fig2} shows that the maximum residual decreases exponentially. 
Although the maximum residual oscillates as a function of the RSCG iteration step $k$, the calculated mean fields are converged, since 
the mean fields are obtained by the sum of the Green's functions with the Matsubara frequencies $\sigma_{j} = i \omega_{j}$. 
We note that the number of the iteration steps in the RSCG method at the site outside the circle is only two. 
At the center of the system $(i_{x},i_{y}) = (L_{x}/2,L_{y}/2)$, the number is about 1000.

\begin{figure}[t]
%%%%--- I comment out figure regions
%\vspace{50mm}
\begin{center}
     \begin{tabular}{p{ 0.8 \columnwidth}} %p{0.5 \columnwidth}}%  p{28mm}}
      \resizebox{0.8 \columnwidth}{!}{\includegraphics{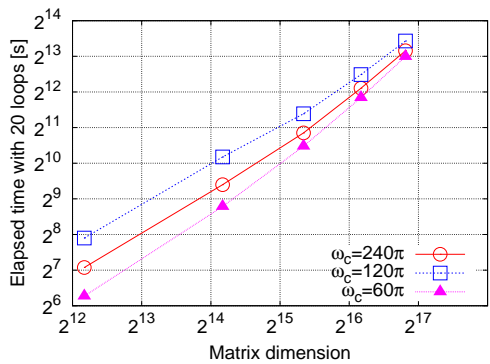}} 
%      &
      %\\ %&
 %     \resizebox{0.8 \columnwidth}{!}{\includegraphics{Fig1b.eps}} 
    \end{tabular}
\end{center}
\caption{
(Color online) Matrix-dimension dependence of the elapsed time with 20 BdG loops. 
The matrix dimension $M$ is defined as $M = 2N^{2}$ in the $d$-wave island with $L_{x} \times L_{y} = N \times N$. 
Other parameters are same in Fig.~\ref{fig:fig1}. 
We use 2304 CPU cores in the ICE X in Japan Atomic Energy Agency. 
%
%The maximum of the residual-vector amplitudes ${\rm Max}\: ||r_{k}(\sigma_{j})||$ and the convergence profile of the $d$-wave order parameter at the center  $(i_{x},i_{y}) = (L_{x}/2,L_{y}/2)$. The parameters are same in Fig.~\ref{fig:fig1}. $\Delta_{k}$ denotes the $d$-wave order parameter at the center on the $k$-th iteration step in the RSCG method. 
\label{fig:ela}
 }
\end{figure}

We show the matrix-dimension dependence $M$ of the elapsed time in the BdG loops. 
Here, the matrix dimension $M$ is defined as $M = 2N^{2}$ in the $d$-wave island with $L_{x} \times L_{y} = N \times N$. 
We measure the elapsed time in 20 times BdG iterations steps with different Matsubara cutoffs $\omega_{\rm c} = \pi T (2 n_{\rm c} +1)$ as shown in Fig.~\ref{fig:ela}. 
We use 2304 CPU cores in the supercomputing system ICE X in Japan Atomic Energy Agency. 
When the matrix dimension is small $(M \ll n_{\rm c})$, the elapsed time grows in  ${\cal O}(M n_{\rm c})$ manner, since the cost calculating the matrix element at each shift point is heavier than that doing the matrix-vector operation and the number of the mean fields at real-space grids is proportional to $M$.  
With increasing the matrix dimension, the elapsed time grows in ${\cal O}(M^{2})$ manner, since the sparse-matrix-vector operation is 
${\cal O}(M)$. 
%
%
%Figure \ref{fig:ela} clearly shows that the elapsed time approximately grows in ${\cal O}(M)$ manner with $n_{c}$. 
%increasing the matrix dimension $M$. 
%Here, the matrix dimension $M$ is defined as $M = 2N$ in the $d$-wave island with $L_{x} \times L_{y} = N \times N$. 
In contrast, the full diagonalization scheme demands ${\cal O}(M^{3})$ cost in the core part of the calculation. 
Thus, we claim that the RSCG method is much faster than that with the full diagonalization method in large systems. 
Note that the converged mean fields with $\omega_{c} = 120\pi$ and $\omega_{c} = 60\pi$ are equivalent to those with $\omega_{c} = 240\pi$. 

\begin{figure}[t]
%%%%--- I comment out figure regions
%\vspace{50mm}
\begin{center}
     \begin{tabular}{p{   \columnwidth}} %p{0.5 \columnwidth}}%  p{28mm}}
%          \begin{tabular}{p{  0.5 \columnwidth}p{ 0.5  \columnwidth}} %p{0.5 \columnwidth}}%  p{28mm}}
      (a)\resizebox{0.8 \columnwidth}{!}{\includegraphics{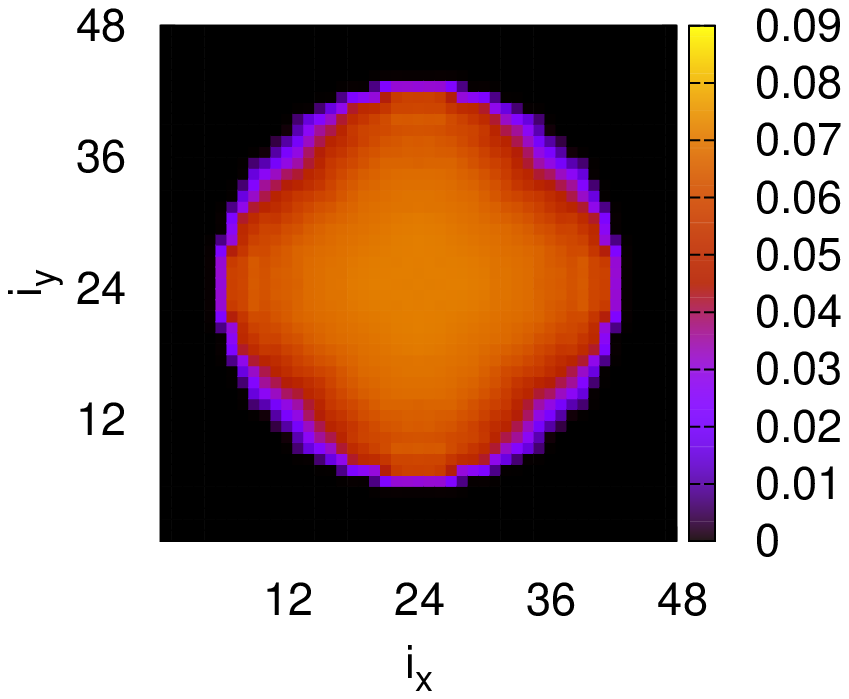}} \\
%      &
      %\\ %&
      (b)\resizebox{0.8 \columnwidth}{!}{\includegraphics{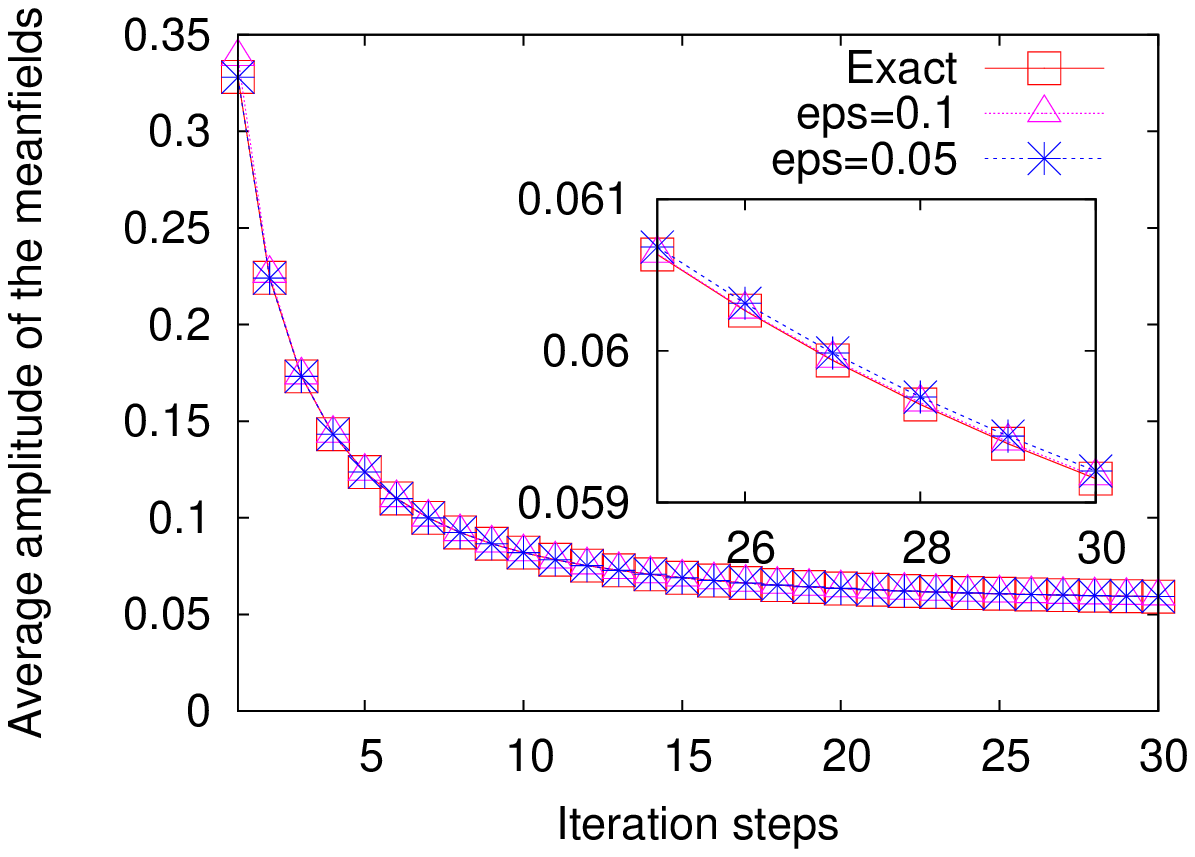}} 
    \end{tabular}
\end{center}
\caption{(Color online) 
(a) Amplitude of the $d$-wave order parameter in the circler nano island and (b) convergence profile of the average amplitudes of the mean fields on the nano island.  
The system size is $L_{x} \times L_{y} = 48 \times 48$. 
The other parameters are same in Fig.~\ref{fig:fig1}.
\label{fig:fig3}
 }
\end{figure}

Let us discuss the accuracy of the converged mean fields. 
To compare with the direct diagonalization of the BdG Hamiltonian, we consider the system with the system size $L_{x}\times L_{y} = 48 \times 48$ whose matrix dimension is 4608. 
The converged $d$-wave order parameter shown in Fig.~\ref{fig:fig3}(a) is smaller than that shown in Fig.~\ref{fig:fig1}(a), since the Andreev bound states at the [110] boundary reduce the $d$-wave order parameter. 
Figure \ref{fig:fig3}(b) shows that this method has a good accuracy to obtain the mean fields. 
The error of the average gap amplitude calculated by the RSCG method with the convergence criterion $0.1$ is $2 \times 10^{-4}$. 
The number of the matrix-vector operations to reach this criterion is 350. 
Here, we define the error as $|\Delta_{\rm exact}-\Delta_{\rm RSCG}|/\Delta_{\rm exact}$ with the average mean field calculated by the direct diagonalization $\Delta_{\rm exact}$ and that calculated by the RSCG method $\Delta_{\rm RSCG}$ at the 30th BdG iteration step.  
In the direct diagonalization, we use Eq.~(\ref{eq:direc}) to calculate the Green's functions. 
The ``eps'' is the accuracy of the simultaneous linear equations (\ref{eq:meanx}), which is defined as  the cutoff value of the maximum of the residual-vector amplitudes ${\rm Max}\: ||r_{k}(\sigma_{j})||$ in Eq.~(\ref{eq:rk}). 
%This means that 500 matrix-vector operations are enough to obtain the accurate mean field at a site. 

%
%
%At the 110 boundaries, 
%The zero energy local density of states shown in  are calculated by the SS method.
%
%
%The number of the iteration steps in the RSCG method at the site outside the circle is only two. 
%
%
%The Andreev bound states at the 110 boundaries\cite{Hu} can be seen clearly, which is similar to the results 
%
%At the center of the system $(i_{x},i_{y}) = (L_{x}/2,L_{y}/2)$, the number is about 2500 as shown in Fig.~\ref{fig:fig2}. 
%Figure \ref{fig:fig2} also shows that the maximum residual decreases exponentially. 
%We should note that Fig.~\ref{fig:fig2} does not indicate the accuracy of the mean field, since the mean field is obtained by the sum of the Green's functions with the energy $\sigma_{j}$. 
%
%
%
%
%
%
%
%

\subsection{Vortex lattice}
\begin{figure}[t]
%%%%--- I comment out figure regions
%\vspace{50mm}
\begin{center}
     \begin{tabular}{p{ 0.8 \columnwidth}} %p{0.5 \columnwidth}}%  p{28mm}}
      \resizebox{0.8 \columnwidth}{!}{\includegraphics{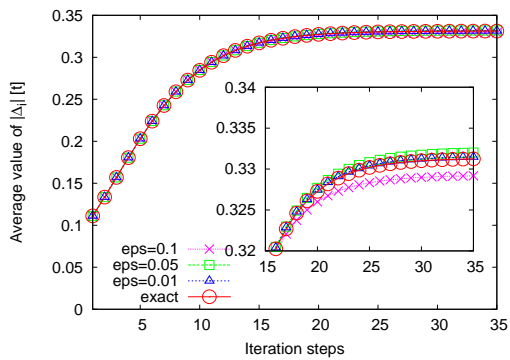}} 
%      &
      %\\ %&
 %     \resizebox{0.8 \columnwidth}{!}{\includegraphics{Fig1b.eps}} 
    \end{tabular}
\end{center}
\caption{(Color online) Average of the gap amplitude in the vortex lattice system in the $s$-wave superconductor. The onsite pairing interaction $U = -2.5t$. The system size is $L_{x} \times L_{y} = 30 \times 30$. Other parameters are same in Fig.~\ref{fig:fig1}.
\label{fig:fig4}
 }
\end{figure}

To investigate the performance of the RSCG method with the complex hermitian Hamiltonian, we consider the vortex lattice system in the $s$-wave superconductor with the  system $L_{x} \times L_{y} = 30 \times 30$ whose matrix dimension is 1800. 
We consider $s$-wave onsite pairing interaction $U_{ijkl} = \delta_{ik} \delta_{jl} \delta_{ij} U$ and we set $U = -2.5t$. 
There are two vortices per unit cell\cite{Evan}. 
The vector potential is considered as a Peierls phase\cite{Nagaid-wave}.    
The other parameters are same in the previous section. 
Figure \ref{fig:fig4} shows that the RSCG method has a good accuracy even in the case with the complex hermitian Hamiltonian.  
The error of the average gap amplitude calculated by the reduced shifted method with the convergence criterion $0.01$ is $7 \times 10^{-4}$. 
%This means that 2000 matrix-vector operations are enough to obtain the accurate mean field at a site in this system. 

\section{Technical Remarks}\label{sec:tech}
In this section, we refer to technical remarks to describe the advantages of the RSCG method. 
These points are useful when actually performing large-scale numerical calculations. 

\subsection{Matrix-vector operations}
The algorithm of the matrix multiply operation $\Vec{y} = \hat{A} \Vec{x}$ is most important part for a fast calculation in the RSCG method. 
Since the target Hamiltonian is generally a sparse matrix, we can choose a fast algorithm optimized for the sparse-matrix vector product among several suggested ones. 
As we pointed out in the case of the Kernel polynomial method\cite{NagaiJPSJ}, we confirm that Compressed Row Storage (CRS) format, which is one of the 
typical storing-schemes for sparse matrices, is quite useful for the present RSCG method. 
The CRS format puts the subsequent nonzeros of the matrix row in contiguous memory locations. 
The algorithm with the CRS format is efficient on scalar processors since it has unit stride access.  

\subsection{Convergence property}
The number of the iteration steps to calculate the Green's functions with a desired accuracy in the RSCG method is determined by the distance from the poles of the Green's functions. 
Generally, the linear equation $Ax = b$ with the matrix $A$ with zero eigenvalues does not have a solution vector $x$, since there is no inverse matrix of $A$. 
Thus, the CG-based method does not converge in this case. 
In terms of the Green's function, nobody can calculate the Green's function at its poles.  
In the case of the BdG equations, 
the numbers of the iteration steps for the Matsubara Green's functions defined on the imaginary axis are smaller than that for the retarded or advanced Green's functions 
defined near the real axis, since the eigenvalues of the BdG Hamiltonian are located on the real axis. 
Thus, if one wants to obtain the physical quantity with a frequency very close to zero eigenvalues after the self-consistent calculation, it is better to use the SS method. 

\subsection{Initial guess of the superconducting order parameter}
It is better to consider the large amplitude of the order parameter as the initial guess.  
In the RSCG method, the seed point is set to the origin of the complex frequency plane.  
The convergence property of the RSCG loop is better when the eigenvalues are far from the origin. 
Putting the large superconducting gap as the initial guess, many eigenvalues are located above the superconducting energy gap.

\subsection{How to treat $d$-wave order parameter}
In the $d$-wave superconductor, there are four nonlocal order parameters as shown in Eq.~(\ref{eq:deld}). 
We point out that the mean fields $\langle c_{i} c_{j} \rangle$ with different $j$ sites and a fixed $i$ site are simultaneously obtained 
by the RSCG method, since the linear equation (\ref{eq:meanx}) at a site $i$ does not depend on $j$.  
In the $d$-wave superconductor, $4 \times 2N$ matrix $V$ is needed.

\subsection{Possible applications}
The RSCG method can treat the continuous model by discretizing differential equations.  
On the other hand, the KPM is not suitable in this case, since a maximum eigenvalue increases with decreasing a discretized mesh size. 
The renormalize factor of the KPM should be large when the discretized mesh size is small so that the energy resolution of the KPM 
decreases with a fixed polynomial cutoff. 

The RSCG method does not require a hermitian or symmetric matrix, since the linear equation with a non-symmetric matrix can be solved by the conjugate-gradient-based method. 
This means that the Green's function with complex frequency poles can be calculated by the RSCG method. 
On the other hand, the Hamiltonian matrix should be hermitian in the KPM. 
For example, 
the Hamiltonian defined on polar coordinates, which is a non-hermitian matrix with real eigenvalues, can be treated by the RSCG method. 
The bosonic Bogoliubov equations with complex frequencies can be also solved by this method.

\subsection{Parallel computation}
It is easy to implement a parallel computation in the RSCG method. 
Equations (\ref{eq:meanx}) with different index $i$ can be solved separately. 
Thus, the separate calculations of $[\hat{\Delta}]_{ij}$ is performed on each CPU core. 
The communication in this case, which is a one-to-all communication, is needed only when updating $[\hat{\Delta}]_{ij}$. 

\subsection{On demand RSCG method}
One can calculate quantities on shift points after calculating a quantity on a seed. 
In the RSCG method, 
the significant reduction of stored memory has a potential to allow us very flexible solution.
By the reduction, whole sequence of the reduced vector $\Sigma_k$ is able to be stored at a memory as well as a storage, {\it e.g.} five thousand iterations for a complex reduced vector $\Sigma_k \in \mathbb{C}^4$ requires just 64 Kbyte.
Since the whole reduced vectors can be constructed by just $\Sigma_k$ and expected $\sigma_j$ with use of tenth to seventeenth line in table \ref{table:RSCG}, they are not necessarily calculated in the sequence on a seed but done after constructing $\Sigma_k$ for {\it on demand} shifts.
The on demand RSCG algorithm is specified in table \ref{table:RSCG2}.
Similar strategy is hardly achieved for usual shifted CG for a large problem, because the number of dimension in a problem is usually large enough, {\it e.g.} five thousand iterations for a full complex residual $r_k \in \mathbb{C}^{36864}$ requires 2.95 Gbyte.
The discrepancy between two problem increase as trivially proportional to number of dimension in expected problem with fixed dimension for the reduced vectors.

\begin{table*}[hbt]
  \caption{On demand reduced-shifted CG for $(\sigma_{j} I + A) x(\sigma_{j}) = b$ with a hermitian matrix $A \in \mathbb{C}^{n \times n}$. $V \in \mathbb{C}^{m \times n}$, $x_{k},r_{k},p_{k} \in \mathbb{C}^{n}$, 
  $\alpha_{k}(\sigma_{j}),\beta_{k}(\sigma_{j}),\rho_{k}(\sigma_{j}),\Xi_{k}(\sigma_{j}),\Pi_{k}(\sigma_{j}),\Sigma_{k}  \in \mathbb{C}^{m}$, and $\alpha_{k},\beta_{k} \in \mathbb{C}$.
  $\Xi_{k}(\sigma_{j}) \equiv V x_{k}(\sigma_{j})$. $V^{\rm T} \equiv (v_{1},v_{2},\cdots,v_{m})$. $v_{i} \in \mathbb{C}^{n}$.
  }
  \label{table:RSCG2}
  \begin{center}
    \begin{tabular}{ c l }
      \hline 
      1.  & Set $x_{0}=0$, $r_{0}=p_{0}=b$, $\alpha_{-1}=1$,$\beta_{-1}=0$ \\
      2.  & Compute $\Sigma_{0} = V b$ \\
      3.  & For $k=0,1,\dots$ until convergence Do: \\
      4.  & ~~~~~~$\alpha_k=(r_{k},r_{k})/(p_k,Ap_k)$ \\
      5.  & ~~~~~~$x_{k+1}=x_{k}+\alpha_k p_k$ \\
      6.  & ~~~~~~$r_{k+1} = r_k - \alpha_k Ap_k$ \\
      7.  & ~~~~~~$\beta_k=(r_{k+1},r_{k+1})/(r_k,r_k)$ \\
      8.  & ~~~~~~$p_{k+1} = r_{k+1} + \beta_k p_k$ \\
      9.  & ~~~~~~Compute $\Sigma_{k+1} = V r_{k+1}$ and write it to a storage\\
      10. & End Do \\
    \hline 
      11. & Input $\sigma_{j}$($j = 1,2,\cdots,N$) \\
      12  & Read $\Sigma_{k+1} = V r_{k+1}$ from the storage\\
      13. & Set $\Xi_{0}(\sigma_{j})=0$, $\Pi_{0}(\sigma_{j})=\Sigma_{0}$, $\rho_{-1}(\sigma_{j})=\rho_{0}(\sigma_{j})=1$ ($j=1,2,\cdots,N$) \\
      14. & For $k=0,1,\dots$ until convergence Do: \\
      15. & ~~~~~~For $j = 1,2,\cdots,N$ Do: \\
      16. & ~~~~~~~~~~~~$\rho_{k+1}(\sigma_{j}) =\frac{\rho_{k}(\sigma_{j})\rho_{k-1}(\sigma_{j}) \alpha_{k-1}}{\rho_{k-1}(\sigma_{j}) \alpha_{k-1}(1+\alpha_{k} \sigma_{j})+\alpha_{k} \beta_{k-1} (\rho_{k-1}(\sigma_{j})-\rho_{k}(\sigma_{j})) }$\\
      17. & ~~~~~~~~~~~~$\alpha_{k}(\sigma_{j})=\frac{\rho_{k+1}(\sigma_{j})}{\rho_{k}(\sigma_{j})} \alpha_{k}$ \\
      18. & ~~~~~~~~~~~~$\Xi_{k+1}(\sigma_{j})=\Xi_{k}(\sigma_{j}) +\alpha_{k}(\sigma_{j}) \Pi_{k}(\sigma_{j})$ \\
      19. & ~~~~~~~~~~~~$\beta_{k}(\sigma_{j})=\left( \frac{\rho_{k+1}(\sigma_{j})}{\rho_{k}(\sigma_{j})} \right)^{2} \beta_{k}$ \\
      20. & ~~~~~~~~~~~~$\Pi_{k+1}(\sigma_{j})=\rho_{k+1}(\sigma_{j})\Sigma_{k+1}+\beta_{k}(\sigma_{j}) \Pi_{k}(\sigma_{j})$ \\
      21. & ~~~~~~End Do \\
      22. & End Do \\
    \hline 
    \end{tabular}
  \end{center}
\end{table*}

\section{Conclusion} \label{sec:con}
In conclusion, we proposed the efficient numerical solver called the RSCG method to calculate a matrix element of a Green's function. 
Our method is applicable to a general dynamical correlation function with given two operators. 
We showed a power of this solver to consider a nano-structured superconductor. 
In this method, the matrix element of the Green's function is calculated separately with a desired accuracy. 
One can use a trivial parallel computations to solve the mean field. 
This method allows us to treat the system with the fabrication potential, where one can not use the kernel-polynomial-based method effectively. 
We showed the $d$-wave nano-island as an example. 
This efficient method can be applied in the various kinds of fields in physics to calculate the resolvent of the Hamiltonian.

\section*{Acknowledgment}

Y.~N.~ would like to acknowledge K. Tanaka, Susumu Yamada and Masahiko Machida for helpful discussions and comments. 
The calculations were performed by the supercomputing system SGI ICE X at the Japan Atomic Energy Agency. 
This study was partially supported by JSPS KAKENHI Grant Number 26800197, 15K00178 and the “Topological Materials Science” (No. JP16H00995) KAKENHI on Innovative Areas from JSPS of Japan. 

\clearpage
%\section*{references}

\clearpage
%\newpage
%\widetext
%\onecolumngrid


\begin{thebibliography}{99}
\bibitem{note}
This trick trivially can be applied to any shifted Krylov solvers which compose only linear operation in the sequence.
\end{thebibliography}
\end{document}